\documentclass[amsmath,amssymb,twocolumn,amsfonts]{revtex4-2}
\usepackage{graphicx}
\usepackage{csquotes}
\usepackage[T1]{fontenc}
\usepackage{upgreek}
\def \beq {\begin{equation}}
\def \eeq {\end{equation}}
\def \ba {\begin{align}}
\def \ea {\end{align}}
\newcommand {\qs} {{\rm Q}_{\rm S}}

\newcommand {\ks} {k_{\rm S}}
\newcommand {\kt} {k_{\rm T}}
\usepackage{bm}
\usepackage{bbm,mathbbol}
\usepackage{braket,bigints}
\usepackage{stmaryrd,mathtools}

\newcommand{\tr}{\rm Tr}

\begin{document}
\title{Approaching the quantum limit of energy resolution in animal magnetoreception}
\author{I. K. Kominis}
\email{ikominis@uoc.gr}
\author{E. Gkoudinakis } 
\affiliation{Department of Physics and Institute of Theoretical and Computational Physics, University of Crete, Heraklion 70013, Greece}
\begin{abstract}
A large number of magnetic sensors, like superconducting quantum interference devices, optical pumping and nitrogen vacancy magnetometers, were shown to satisfy the energy resolution limit. This limit states that the magnetic sensitivity of the sensor, when translated into a product of energy with time, is bounded below by Planck's constant, $\hbar$. This bound implies a fundamental limitation as to what can be achieved in magnetic sensing. Here we explore biological magnetometers, in particular three magnetoreception mechanisms thought to underly animals' geomagnetic field sensing: the radical-pair, the magnetite and the MagR mechanism. We address the question of how close these mechanisms approach the energy resolution limit. At the quantitative level, the utility of the energy resolution limit is that it informs the workings of magnetic sensing in model-independent ways, and thus can provide subtle  consistency checks for theoretical models and estimated or measured parameter values, particularly needed in complex biological systems. At the qualitative level, the closer the energy resolution is to $\hbar$, the more \enquote{quantum} is the sensor. This offers an alternative route towards understanding the quantum biology of magnetoreception. It also quantifies the room for improvement, illuminating what Nature has achieved, and stimulating the engineering of biomimetic sensors exceeding Nature's magnetic sensing performance.
\end{abstract}
\maketitle 
\section{Introduction}
Animal magnetoreception is a major and strongly interdisciplinary scientific puzzle \cite{Keeton,Kalmijn,Lohmann1,Wiltschko2005,Lohmann2,Mouritsen2018,JXu,Bellinger,Xie2022}. Quoting a 2010 paper \cite{Lohmann3}, \enquote{..the solution to the magnetoreception mystery will almost certainly come from a fascinating interplay of biology, chemistry and physics}. The quote is as valid today as it was in 2010, since science still awaits the full solution to this challenging problem of sensory biology. 

Numerous behavioural experiments have established the ability of many species to sense the geomagnetic field, yet the mechanisms behind biological magnetic sensing, and their working at the physical, chemical and physiological level, are still being debated. There is evidence that migratory birds \cite{bird1,bird2,bird3,bird4,bird5}, fruit flies \cite{flies1,flies2,flies3,flies4,flies5,flies6,flies7,flies8,flies9}, butterflies \cite{butterfly1,butterfly2,butterfly3}, turtles \cite{turtle1,turtle3,turtle2}, sharks \cite{shark1,shark2,shark3}, eels \cite{eel1,eel2,eel3,eel4,eel5}, salmon \cite{salmon1,salmon2,salmon3}, are among the many species having magnetic sensing capabilities, either at the level of a compass for acquiring directional information towards long-distance navigation, or in some form of a magnetic map for positioning, or both. Three mechanisms have so far been prevalent in the discussion of biological magnetoreception, the radical-pair \cite{rpm1,rpm2,rpm3,rpm4,rpm5}, the magnetite \cite{magn1,magn2,magn3,magn4,magn5}, and the induction mechanism \cite{ind1,ind2,ind3,ind4}, while a mechanism synthesizing the first two, the MagR protein complex, was relatively recently proposed \cite{Xie1}.

Sensing the magnetic field is by all means a physical process. In this respect, biological magnetometers must have quite some common ground with man-made magnetometers, which in recent decades have made spectacular progress towards highly sophisticated devices. Magnetometers largely based on classical physics have been around for several decades \cite{fluxgate1,fluxgate2,Ripka}. However, significant progress in magnetic sensitivity was driven by modern quantum technology \cite{Dowling,Deutsch}, a major thrust of which is quantum sensing \cite{Degen} of magnetic fields. 

The reason behind this progress is that at the quantum level, measuring a magnetic field is equivalent to measuring energy. The precision of energy measurements depends on the lifetime of \enquote{quantumness}, which reflects how well a \enquote{fragile} quantum system is isolated from its perturbing environment. Modern quantum experiments have pushed this isolation to new extremes, thus leading to sensitive magnetic sensing technologies. Among those are superconducting quantum interference devices  \cite{squid1,squid2,squid3}, optical pumping magnetometers \cite{opm1,opm2} , diamond sensors \cite{dm1,dm2,dm3}, Bose-Einstein condensates \cite{bec1,bec2,bec3}. 
\begin{figure*}[ht]
\begin{center}
\includegraphics[width=17.5 cm]{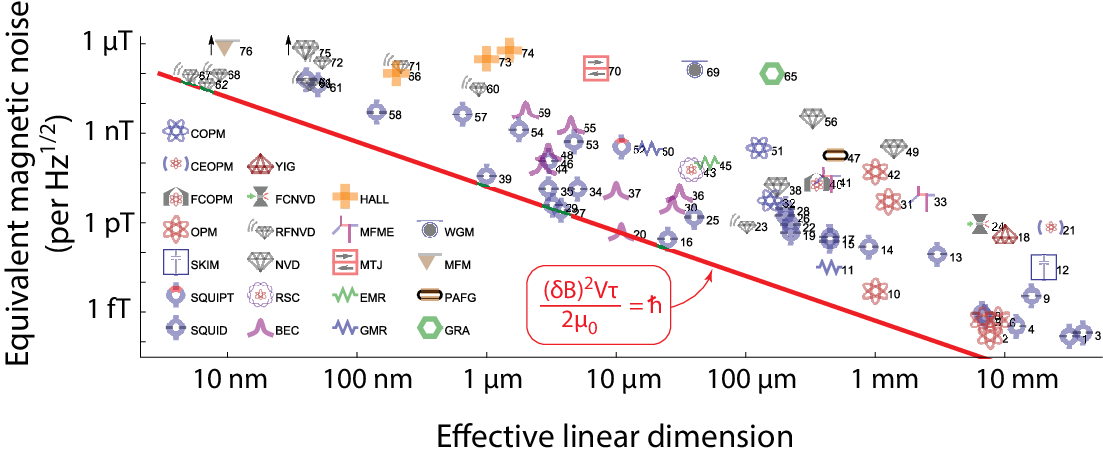}
\caption{Energy resolution limit in magnetic sensing \cite{MitchellRMP}. Magnetic sensitivity versus sensor's linear dimension for tens of realizations of several magnetic sensing technologies, like atomic magnetometers, SQUIDs, diamond and Hall effects sensors, and more. The red solid line is defined by the energy resolution limit, that is, the bound $(\widetilde{\delta B})^2V/2\mu_0\geq\hbar$, where $\widetilde{\delta B}$ is the magnetic sensitivity (or equivalently magnetic noise) in units of ${\rm T/\sqrt{Hz}}$, $V$ is the volume of the sensor (the third power of the linear dimension), and $\mu_0$ vacuum's magnetic permeability constant. If the measurement time is $\tau$, and the error in the magnetic field estimate obtained within the measurement time is $\delta B$, then $\widetilde{\delta B}=\delta B\sqrt{\tau}$. Figure reproduced from \cite{MitchellRMP} with permission of the American Physical Society.} 
\label{Mitchell_ERL}
\end{center}
\end{figure*}

A major goal of magnetometry is to understand how the physical principles underlying each technology limit the magnetic sensitivity and other figures of merit of the sensor. Quantum physics indeed puts stringent limits to what can be achieved with any kind of sensor, although many times it is challenging to precisely define those limits. Nevertheless, it is broadly accepted that such fundamental measurement limits do exist, and research towards understanding and utilizing them to design new magnetic sensing applications is overly active \cite{meg1,meg2,meg3,meg4,meg5,mcg1,mcg2,materials1,materials2,Aslam,Bao}.

Recently, another approach gave a more holistic and {\it technology-independent} perspective on magnetic sensing based on the so-called energy resolution limit (ERL) \cite{MitchellRMP}. The energy resolution (ER), more precisely called energy resolution per bandwidth, is a quantity composed of the uncertainty in the magnetic field estimate performed by the sensor, the sensitive volume, and the measurement time. This quantity has unit of [energy][time], which is the same unit as Planck's constant, $\hbar$. 
It turns out  \cite{MitchellRMP} that numerous magnetic sensing realizations spanning many different technologies have energy resolution larger than $\hbar$. 

Here we use the ERL as a guiding principle for understanding biological magnetoreception using just three basic quantities: sensitivity $\delta B$, volume $V$, and measurement time $\tau$. Biological magnetic sensors should also satisfy the ERL, because after all, biological systems operate under the laws of physics \cite{Meister}. The ERL can offer multiple insights into animal magnetoreception. 

Qualitatively, the closer to $\hbar$ the numerical value of the ER is for any given sensor, the more \enquote{quantum} can the sensor be characterized. This offers an alternative perspective in the study of quantum biology of magnetoreception \cite{qb1,qb2,qb3,qb4,qb5,qb6,qb7,qb8,qb9,qb10}. This discussion is not about a choice between \enquote{black} or \enquote{white}, since \enquote{quantumness} can now be quantified, and spans a continuum. Indeed, quantum information science has introduced quantifiers for quantum coherence \cite{coh} and entanglement \cite{ent}, and some have been applied to quantifying \enquote{quantumness} of biological magnetoreception \cite{kom1,kom2}. The ER provides yet another quantifier, this time in units of $\hbar$. For example, if it is found that some biological magentic sensor has ER e.g. $10^7\hbar$, and thus is far from the quantum limit, this finding is still quite informative, since it implies room for $10^7$ improvement. Even if Nature did not have the chance to realize such an improvement, it could in principle be engineered with biomimetic sensors \cite{biomim1,biomim2,biomim3}. 

Quantitatively, the ERL provides a strong and subtle constraint of basic and measurable parameters ($\delta B$, $V$, $\tau$). This constraint is particularly helpful for understanding  biological magnetoreceptors, which unavoidably face the complexities and uncertainties of biological systems, as opposed to precisely controlled quantum magnetometers in the laboratory. As will be outlined in the paper, some biological ERs cannot be attributed a specific value like the ERs of man-made sensors, but rather, a range of possible values, stemming from those uncertainties. {\it Conversely, it will be seen that the ERL narrows down those uncertainties}. For example, we here obtain a precise lower bound for the number of cryptochromes or MagR complexes that are required to obtain a given magnetic sensitivity within a given measurement time.

The outline of this paper is the following. In Sec. II we make some introductory comments about the energy resolution limit aimed at an interdisciplinary readership. We then apply the ERL to the radical-pair (Sec. III), the magnetite (Sec. IV), and the MagR mechanism (Sec. V). In Sec. VI we present the results in a global ERL map. We conclude in Sec. VII.
\section{Introductory comments about the energy resolution limit}
The authors in \cite{MitchellRMP} analyzed the magnetometry literature and observed that tens of different realizations of several magnetic sensing technologies satisfy a bound, which reads
\beq
{\rm ER}\equiv{{(\delta B)^2}\over {2\mu_0}}V\tau\gtrapprox\hbar\label{ERL}
\eeq
The left-hand-side defines the energy resolution of the sensor, and the right-hand side is Planck's constant $\hbar\approx 10^{-34}~{\rm Js}$, which constant pervades quantum phenomena. 

In this expression, $\delta B$ is the uncertainty in the sensor's magnetic field estimate, $V$ the sensitive volume, and $\tau$ the measurement time. The constant $\mu_0=4\pi\times 10^{-7}~{\rm Tm/A}$ is the magnetic permeability of  vacuum. The name \enquote{energy resolution} derives from the following. From classical electromagnetism we know that $B^2/2\mu_0$ is the magnetic energy density (unit ${\rm J/m^3}$) due to the presence of a magnetic field $B$ in space. For a homogeneous magnetic field, the total energy within some volume $V$ is then $B^2V/2\mu_0$. The ERL involves the variance $(\delta B)^2$ instead of $B^2$, therefore one can loosely think of it as the minimum magnetic energy measurable within the sensor volume during the time available for the measurement. The right-hand-side of \eqref{ERL} is approximately equal to $\hbar$, thus it is tempting to guess that some basic quantum principle underlies the ERL. 

Here we will not address the intricacies of the quantum physics of the ERL, but for completeness we note that (i) such a principle was recently proposed to derive from quantum thermodynamics \cite{KominisERL}, and (ii) there are claims for magnetic sensors having an ER below $\hbar$ \cite{belowERL1,belowERL2,belowERL3}. Our understanding \cite{KominisERL} is that such claims involve non-trivial quantum enhancements that one would hardly expect to find in biological systems, even in light of quantum biological effects \cite{KominisReview}. Given that a vast number of technologies do satisfy the ERL \cite{MitchellRMP}, we here work under the premise that the same is the case for biological magnetometers. 
\begin{figure*}[t]
\begin{center}
\includegraphics[width=17.5cm]{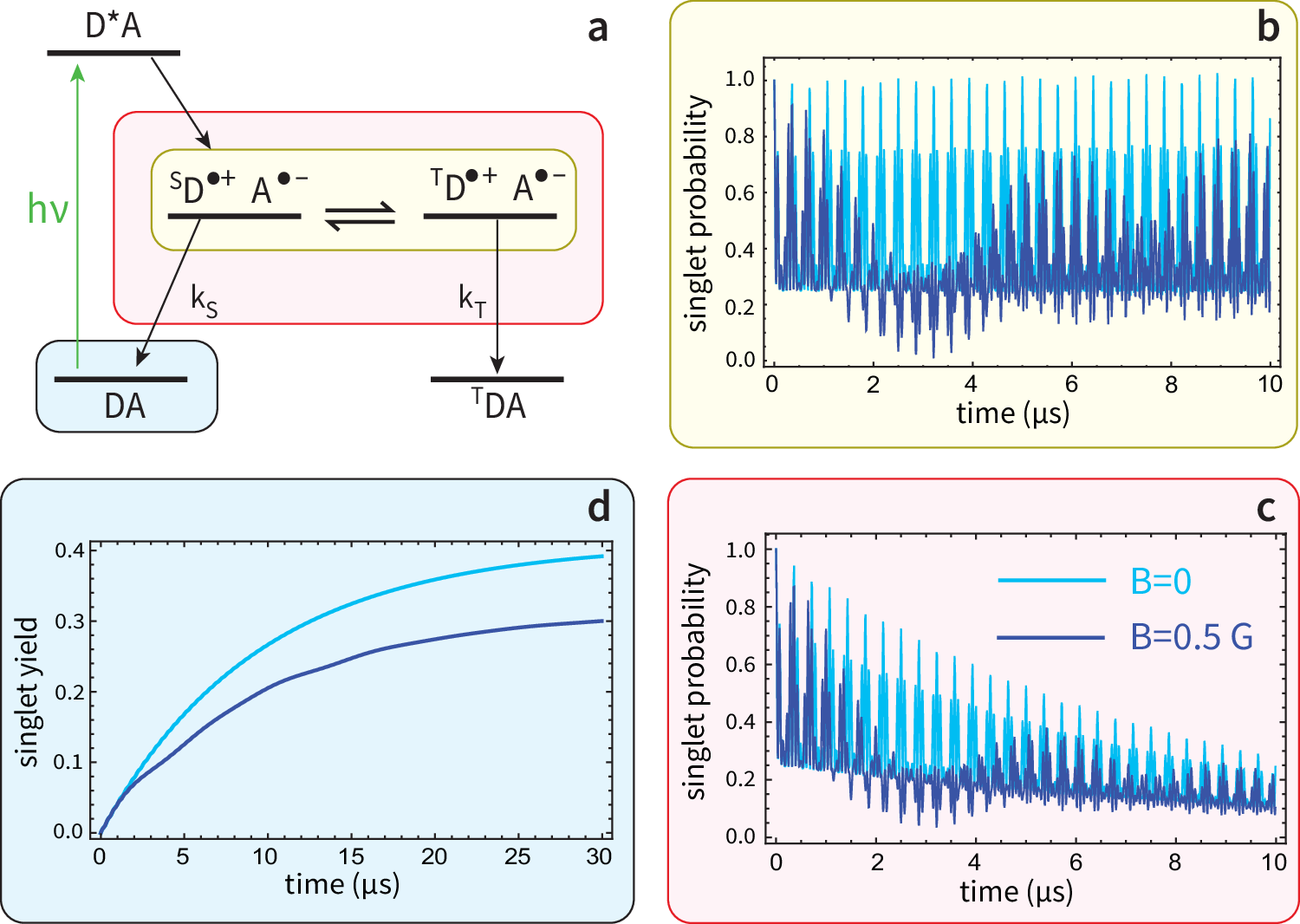}
\caption{Radical-pair mechanism and example of spin dynamics. (a) A donor-acceptor dyad, DA, is photo-excited. The photo-excited state ${\rm D^*A}$, leads by electron transfer to the singlet radical-pair, ${\rm ^SD^{\bullet +}A^{\bullet -}}$. Intra-molecule magnetic interactions, including the Zeeman interaction of the two unpaired electrons with the external magnetic field, drive singlet-triplet oscillations, i.e. a coherent interconversion ${\rm ^SD^{\bullet +}A^{\bullet -}}\leftrightarrow {\rm ^TD^{\bullet +}A^{\bullet -}}$. The inverse electron transfer from A back to D leads to the neutral recombination products through two recombination channels, the singlet (rate $\ks$) producing ${\rm DA}$, and the triplet (rate $\kt$) producing ${\rm ^TDA}$. (b) Example of singlet-triplet oscillations with no recombination, for two different magnetic fields, $B=0$ and $B=0.5~{\rm G}$. The y-axis is the singlet probability, given by $\tr\{\rho_t\qs\}$, where $\rho_t$ is the radical-pair spin density matrix at time $t$, and ${\rm Q_S}$ the singlet projection operator. The model considered here is a radical-pair with two spin-1/2 nuclei, one at D and one at A, the density matrix in this case having dimension 16. The Hamiltonian driving the spin motion is ${\cal H}=B(s_{Dz}+s_{Az})+a_D\mathbf{s}_D\cdot\mathbf{I}_D+a_A\mathbf{s}_A\cdot\mathbf{I}_A$, where $\mathbf{s}_D$ and $\mathbf{s}_A$ are the electron spins of D and A, $\mathbf{I}_D$ and  $\mathbf{I}_A$ the nuclear spins of D and A, and $a_D$ and $a_A$ the respective hyperfine couplings, here considered isotropic. The dynamics are described by the unitary evolution $d\rho/dt=-i[{\cal H},\rho]$. Here $a_D=10~{\rm G}$ and $a_A=1~{\rm G}$. (c) When the recombination channels are taken into account, the singlet probability decays, since population leaves the radical-pairs (i.e. $\tr\{\rho\}$ decreases) and grows in the neutral products, DA and ${\rm ^TDA}$. The dynamics are now described by a master equation $d\rho/dt=-i[{\cal H},\rho]-k(\qs\rho+\rho\qs-2\qs\rho\qs)-k\rho$, where the second term produces singlet-triplet dephasing, and the third term (called reaction term) accounts for recombination with $k_S=k_T=k$. The lifetime of the reaction is defined as $\tau=1/k$. Here $\tau=10~{\rm \upmu s}$. (d) The singlet yield as a function of time, given by $\int_0^tkdt'\tr\{\rho_{t'}\qs\}$, is seen to grow with time as more and more radical-pairs recombine. The final yield, obtained for $t\gg\tau$, is a physiological observable seen to depend on $B$, and thus it can convey the information on magnetic field changes.}
\label{rmp}
\end{center}
\end{figure*}

In Fig. \ref{Mitchell_ERL} we reproduce the results of \cite{MitchellRMP}, displaying the quantity $\widetilde{\delta B}=\delta B\sqrt{\tau}$ (having unit ${\rm T/\sqrt{Hz}}$) versus the linear dimension $L$ of the sensor. This is connected to the volume entering \eqref{ERL} by $V=L^3$. The spectral magnetic sensitivity $\widetilde{\delta B}$ describes how the measurement noise is distributed over the measurement bandwidth, given by $1/\tau$. The utility of $\widetilde{\delta B}$ is that we can produce a common figure for $\widetilde{\delta B}$ versus $L$ for a number of sensors having different measurement times $\tau$. It is seen in Fig. \ref{ERL} that all sensors examined have $\widetilde{\delta B}\geq\sqrt{2\mu_0\hbar/L^3}$, which is another way to express the bound \eqref{ERL}.

In the next three sections we analyze the ER of the radical-pair, the magnetite, and the MagR mechanism. We do not further discuss the induction mechanism, since the relevant ER stands at the level of $10^{26}\hbar$, rendering the mechanism purely classical, and thus far from any potential biomimetic advancement towards the quantum limit. In the beginning of each of the following three sections we briefly recapitulate the workings of the relevant mechanism. To establish the ER, we extract from the literature the magnetic sensitivity $\delta B$, the sensor volume $V$, and the measurement time $\tau$. At the end, we produce a similar map of the global results like the one in Fig. \ref{Mitchell_ERL}. 
\section{Radical-pair Mechanism}
The radical-pair mechanism can provide both a scalar measurement of the magnetic field amplitude, usually referred to as the \enquote{magnetic field effect} \cite{mfe1,mfe2,mfe3,mfe4,mfe5}, as well as directional information. The former might be used in a magnetic map \cite{Mouritsen2018}, while the latter in a compass \cite{compass1,compass2,compass3,compass4}. The experimental evidence supporting radical-pair magnetoreception, reviewed e.g. in \cite{Mouritsen2018}, involves disorientation by radiofrequency magnetic fields, localization in the retina of cryptochrome proteins hosting the radical pairs, and co-localization of cryptochrome with neuronal activity markers \cite{rf1,rf2,rf3,rf4,loc1,loc2,loc3,coloc1}. 

Within the perspective of the ERL we focus on estimating the magnetic field amplitude. This does not limit the utility of our results, since as stated in \cite{Mouritsen2018}, \enquote{map and compass cues might not always be as separable as previously thought}. Information on the magnetic field change is reflected in the change in concentration of radical-pair reaction products. We remind the reader that such reactions involve a coherent spin motion, which is influenced by the magnetic field. This spin motion is driven by the intra-molecular magnetic fields produced by the hyperfine couplings of the two unpaired electrons of the donor and acceptor molecule forming the radical-pair with the magnetic nuclei of the respective molecule. The radical-pair is created by photo-excitation of the neutral precursor. The mechanism is depicted in Fig. \ref{rmp}a, which shows the photo-excitation step of the donor-acceptor molecular dyad, DA, the electron transfer creating the radical-pair state ${\rm D^{\bullet +}A^{\bullet -}}$ in the singlet state of the two unpaired electrons, the coherent spin motion ${\rm ^SD^{\bullet +}A^{\bullet -}}\leftrightarrow {\rm ^TD^{\bullet +}A^{\bullet -}}$ interchanging singlet with triplet radical-pair states, and the inverse electron transfer leading to the singlet (DA) and triplet (${\rm ^TDA}$) reaction products. 

The signalling mechanism conveying the change in reaction yields to higher levels of neurophysiological processing is still unknown \cite{Weaver,kom2}. Therefore, as magnetic-dependent observable one usually considers the relative proportion of singlet versus triplet reaction yields. The singlet (triplet) yield is found by integrating the singlet (triplet) probability of the radical-pair state times the probability for singlet (triplet) recombination, which for a time interval $dt$ is given by $\ks dt$ ($\kt dt$). The singlet (triplet) probability of the radical-pair state is theoretically calculated from the quantum state evolution of the radical-pair spin state (see caption of Fig. 2). The rates $\ks$ and $\kt$ are characteristic parameters of the specific radical-pair considered. In general they are different, however, in many simulations they are taken equal, $k_S=k_T\equiv k$. In this case, also considered herein, $\tau=1/k$ is the reaction time, {\it which will also serve as the measurement time} in our ER considerations.

For completeness, in Fig. \ref{rmp}b we depict an example of the coherent spin motion ${\rm ^SD^{\bullet +}A^{\bullet -}}\leftrightarrow {\rm ^TD^{\bullet +}A^{\bullet -}}$ for an infinite radical-pair lifetime, i.e. $k=0$. We plot the singlet probability as a function of time for two different magnetic fields, to show how the field modulates the singlet-triplet mixing. In Fig. \ref{rmp}c we \enquote{turn on} the recombination channels, so in this case we observe the decay of the singlet probability, since radical-pairs recombine and their population tends to zero, while the population of the reaction products increases. This is shown in Fig. \ref{rmp}d for the singlet reaction yield, which is seen to convey information on the magnetic field. Indeed, for a change from $B=0$ to $B=0.5~{\rm G}$, the singlet reaction yield at the end of the reaction ($t\gg\tau$) changes by 10\% (in absolute terms).
\subsection{Energy resolution of synthesized radical-pairs probed with laser spectroscopy}
The authors in \cite{Maeda2008} synthesized a radical-pair model using a carotenoid-porphyrin-fullerene triad (CPF). The porphyrin serves as an intermediate cation in the electron-transfer process, whereas the actual radical-pair is ${\rm C^{\bullet +}PF^{\bullet -}}$, i.e. the carotenoid is the radical cation, and the fullerene is the radical anion. The authors observed a clear magnetic field effect in the radical-pair reaction at earth's magnetic field, albeit at 113 K, i.e. much cooler than the physiological animal temperature.  Nevertheless, these data are perfectly suitable to apply the ERL. The authors measured the transient absorption of the radical-pair state at $39~{\rm \mu T}$ and  $49~{\rm \mu T}$. From the noise in their data (Fig. 2c of \cite{Maeda2008}) one can estimate the magnetic sensitivity to be $\delta B\approx 2.5~{\rm \mu T}$. The measurement time was $\tau\approx 1~{\rm \mu s}$, given by the duration of the magnetic-field effect. Finally, the sensing volume of the $100~{\rm \mu M}$ CPF solution that was probed by the laser measuring the transient absorption of ${\rm C^{\bullet +}PF^{\bullet -}}$ was $V\approx 2\times 10^{-8}~{\rm m^3}$ \cite{Kerpal2019}. Thus in this case, it is ${\rm ER}=(\delta B)^2V\tau/2\mu_0=5\times 10^{14}\hbar$. 

The main reason for the large value of this ER is the large value of $\delta B$ deriving from the specific measurement scheme. It is known from quantum magnetometers working with atomic vapors \cite{BR} that the number of atoms participating in the measurement is crucial. The number density of CPF molecules, given their $100~{\rm \mu M}$ concentration, is $6\times 10^{16}~{\rm cm}^{-3}$, two orders of magnitude higher than the atom number density of typical optical pumping magnetometers \cite{BR}, which deliver magnetic sensitivity at the level of 1 fT, with a corresponding ER at the level of $10\hbar$. Thus, $\delta B$ of CPF is not limited by the number of CPF molecules (or the photo-excited fraction thereof). Similarly, atomic magnetometers have about 3-4 orders of magnitude longer relaxation time, so again, $\tau$ is not the limiting factor here. Instead, it appears that the large value of the CPF energy resolution is dominated by the limited sensitivity of the specific measurement scheme. 

Indeed, the magnetic field effect is measured by detecting an optical absorption exciting the radical-pair state into higher electronically excited states, without any discrimination between singlet or triplet radical-pairs. That is, the magnetic field effect probes the total population of the radical-pair state, given by $\tr\{\rho\}$, where $\rho$ is the radical-pair density matrix. It can be seen from the master equation evolving $\rho$ (see caption of Fig. \ref{rmp}) that when calculating $d\tr\{\rho\}/dt$, the Hamiltonian term that contains the magnetic field drops out. When the recombination rates are equal, as in the example of Fig. \ref{rmp}, it is $d\tr\{\rho\}/dt=-k\tr\{\rho\}$, and the sensitivity of $\tr\{\rho\}$ to the magnetic field is identically zero. The authors in \cite{Maeda2008} observed a magnetic field effect because for ${\rm C^{\bullet +}PF^{\bullet -}}$ it is $\ks\neq\kt$, and the reaction terms for unequal recombination rates indirectly produce a magnetic-field dependence of $\tr\{\rho\}$. Still, the magnetic sensitivity when measuring $\tr\{\rho\}$ is quite lower than when detecting e.g. the singlet product yield, as magnetoreceptive organisms are supposed to do physiologically. Additionally, molecular excitation is not as sharp as the laser probing of atoms, hence the measurement noise appears to be significant. 

Overall, the specific ER over-satisfies the ERL bound. The values for $\widetilde{\delta B}=\delta B\sqrt{\tau}$ and the spatial dimension $L$ that we will use for CPF magnetoreception in our global results of Sec. VII are $\widetilde{\delta B}\approx 2.5~{\rm nT/\sqrt{Hz}}$ and $L\approx 3~{\rm mm}$.
\subsection{What is $\delta B$ in vivo?}
For the rest of the discussion in this and the following sections, which has to do with animal magnetoreception, we will use a common value for the magnetic sensitivity $\delta B$, which we will now infer from the literature. Then, in each of the subsequent models we will additionally obtain the measurement time $\tau$ and the sensitive volume $V$ in order to arrive at the corresponding ER.

Several studies of the radical-pair mechanism involve the compass operation resulting from anisotropic hyperfine couplings in the radical-pair \cite{aniso1,aniso2,aniso3}, and leading to $\phi$-dependent reaction yields, where $\phi$ is the angle between the magnetic field and some particular axis defined by a hyperfine tensor. Here we do not consider the compass operation, but the estimate of the magnetic field amplitude. For an order-of-magnitude estimate with a hand-waving approach, we could argue that if the mechanism working as a compass has angular precision $\delta\phi$, then one could expect a magnetic field sensitivity $\delta B\approx B\delta\phi$, where $B$ is the background field. On average, the geomagnetic field amplitude is $B\approx 45~{\rm \mu T}$. The angular precision of the compass has been estimated  \cite{rpm5,Ren} as $\delta\phi\approx 1^\circ$. Thus $\delta B\approx 0.8~{\rm \mu T}$. 

From another perspective, it is known from tracking migratory birds that the position accuracy of homing attributed to long-range navigational cues is around 50 km \cite{Mouritsen2018,Komolkin}. The magnitude of the geomagnetic field changes by about $3~{\rm nT/km}$ in the north-south direction \cite{Mouritsen2018}, thus over 50 km the change will be $\delta B\approx 0.15~{\rm \mu T}$. 

The authors in \cite{Lohmann2022} mention that magnetic anomalies of earth's field, known to be mostly at the level of 1\%, could offer navigational cues. This argument would imply $\delta B=0.45~{\rm \mu T}$. Finally, the authors in \cite{Semm} use electrophysiological recordings in the trigeminal system of birds and find a $0.2~{\rm \mu T}$ sensitivity, while the authors in \cite{Lohmann2024} suggest sensitivities down to the 20 nT level. 

The aforementioned numbers involve numerous observations with multiple species, therefore it is virtually impossible to add an error bar weighting the above estimates. Hence we here average all of the above estimates and set $\delta B=0.3~{\rm \mu T}$.
\subsection{Energy resolution in vivo: cryptochrome}
\begin{figure}[t!]
\begin{center}
\includegraphics[width=8.5 cm]{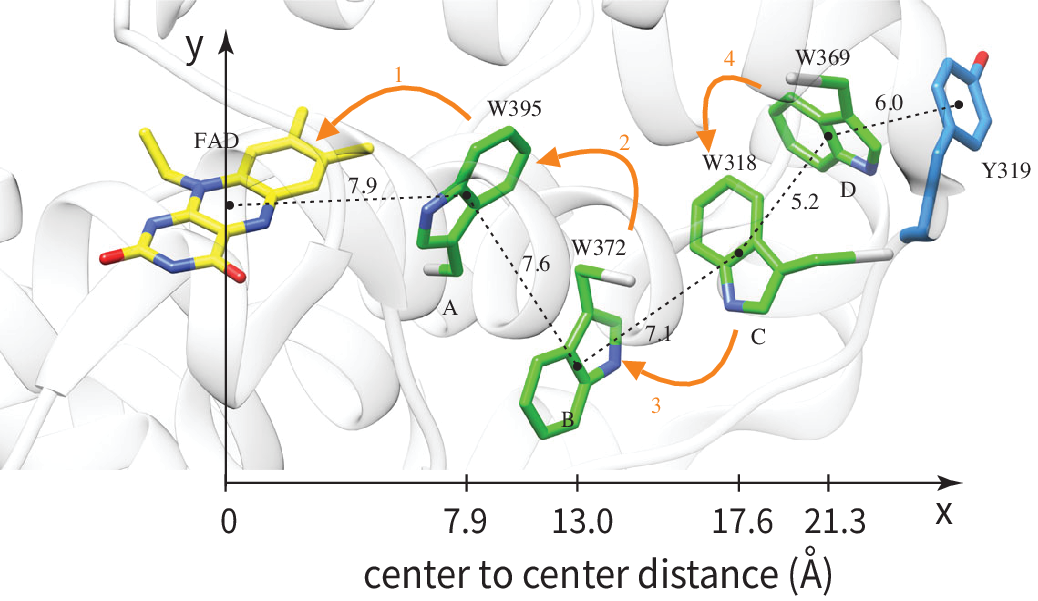}
\caption{Cryptochrome protein (shaded grey background) is the host of the ${\rm FAD^{\bullet -}Tr^{\bullet +}}$ radical-pair working as a magnetic sensor. Center-to-center distances between FAD and tryptophans are known and provide the overall molecular volume of the sensor, $V_{\rm rp}=1.4\times10^{-27}~{\rm m}^3$. We estimate the length of the sensor from the FAD-W369 distance ($21.3~{\rm \mathring{A}}$) and the width from the W372-W369 distance projected on the $y$-axis ($8~{\rm \mathring{A}}$). Figure reproduced from \cite{Wong} under the Creative Commons Attribution License.} 
\label{cry}
\end{center}
\end{figure}

A prevalent model \cite{Schulten2012,Maeda2012,Schulten2014,Nohr2016,Michael,Solovyov,Zoltowski,Hochstoeger,Parico,Wong,Chandrasekaran,Frederiksen,Deviers,Schuhmann} for radical-pair based magnetoreception has been the radical-pair reaction following electron transfer from a triad of tryptophan residues (Tr) to flavin adenine dinucleotide (FAD), with the radical-pair ${\rm FAD^{\bullet -}Tr^{\bullet +}}$ embedded in the protein cryptochrome (Cry), in particular Cry4. This protein is expressed in the retina of the European robin and exhibits strong binding with flavin \cite{Mouritsen2018,Mouritsen2020,Gunther,Muheim}. Recently a tetrad of tryptophans was considered \cite{Wong}, and we will here use this model, depicted in Fig. \ref{cry}. In the same figure we depict the known distances between the radicals, from which we can estimate the volume of the sensor as $V_{\rm rp}=2.1~{\rm nm}\times (0.8~{\rm nm})^2=1.4\times10^{-27}~{\rm m}^3$. However, one can argue that the volume to be used in the estimate of the ER is the Cry volume, which is \cite{Zoltowski} $V_{\rm cry}\approx 5\times 10^{-25}~{\rm m}^3$. We will here use as sensitive volume the radical-pair volume, i.e. set $V=V_{\rm rp}$, and elaborate further on this issue in Sec. III.D. 

Lastly, we need to estimate the measurement time $\tau$, given by the lifetime of the ${\rm FAD^{\bullet -}Tr^{\bullet +}}$ radical-pair. The lifetime of ${\rm FAD^{\bullet -}Tr^{\bullet +}}$ in vivo is not yet known. In vitro measurements at $1~{\rm ^\circ C}$ report $6~{\rm \upmu s}$ \cite{Biskup,Kattnig}. As suggested in \cite{Fay}, a lower bound on $\tau$ is given by the electron spin precession time in earth's field, which is $0.7~\upmu{\rm s}$. The authors in \cite{Worster} find an operational window of the compass for $\tau=10~\upmu {\rm s}$, given the relaxation produced by rotational motion of the radical-pair, in particular FAD, and taking into account the sharp \enquote{needle} described in \cite{rpm5}. The latter is an impressively sharp feature appearing in the angular dependence of the reaction yields for lifetimes larger than $\tau=5~\upmu {\rm s}$, when the realistic hyperfine tensors for ${\rm FAD^{\bullet -}Tr^{\bullet +}}$ are used. If we allow ourselves to hypothesize, it seems unlikely that Nature did not evolve to take advantage of this feature. Based on all of the above, it appears reasonable to set $\tau=10~\upmu {\rm s}$. Towards a more general treatment we introduce the dimensionless factor ${\tilde \tau}={\tau\over {10~\upmu {\rm s}}}$, and let $\tau$ vary between  $1~\upmu {\rm s}$ and $100 ~\upmu {\rm s}$, thus ${\tilde \tau}$ ranges from 0.1 to 10.

Putting it altogether, it follows that ${\rm ER}=4.8\times 10^{-6}{\tilde \tau}\hbar$. Is the ERL violated? Not at all. The previous result shows that the signal from at least $N_{\rm cry}^0(\tilde{\tau})=1/(4.8\times 10^{-6}{\tilde \tau})\approx 2\times 10^5/\tilde{\tau}$ Cry proteins must be physiologically integrated in order to obtain the magnetic sensitivity $\delta B$ within the time $10{\tilde \tau}~\upmu {\rm s}$. {\it This is exactly the utility of the ERL, it informs the workings of this mechanism in model-independent ways}, using as input a few basic parameters. If, for example, it is found that indeed there are only $N_{\rm cry}^0(\tilde{\tau})$ proteins contributing, then Cry magnetoreception will be working right at the quantum limit. On the other hand, if it is found that there are e.g. ten times as many proteins contributing, then Cry magnetoreception will be working at the level of $10\hbar$, and there is in principle room for a ten-fold improvement, were one to mimic Nature.

Overall, the values for $\widetilde{\delta B}=\delta B\sqrt{\tau}$ and the spatial dimension $L$ that we will use for Cry magnetoreception in our global results of Sec. VII are $\widetilde{\delta B}\approx \sqrt{{\tilde \tau}}~{\rm nT/\sqrt{Hz}}$ and $L=(N_{\rm cry}V)^{1/3}\geq 65~{\rm nm}/\tilde{\tau}^{1/3}$, where $N_{\rm cry}$ is the number of Cry proteins contributing to the estimate of the magnetic field, and necessarily, it is $N_{\rm cry}\geq N_{\rm cry}^0(\tilde{\tau})$.
\subsection{Comment on the sensor volume}
Defining the sensor volume of the radical-pair magnetoreceptor is not as straightforward as it might seem, we therefore will elaborate on this issue from a number of different perspectives. 

\begin{figure}[t!]
\begin{center}
\includegraphics[width=7.5 cm]{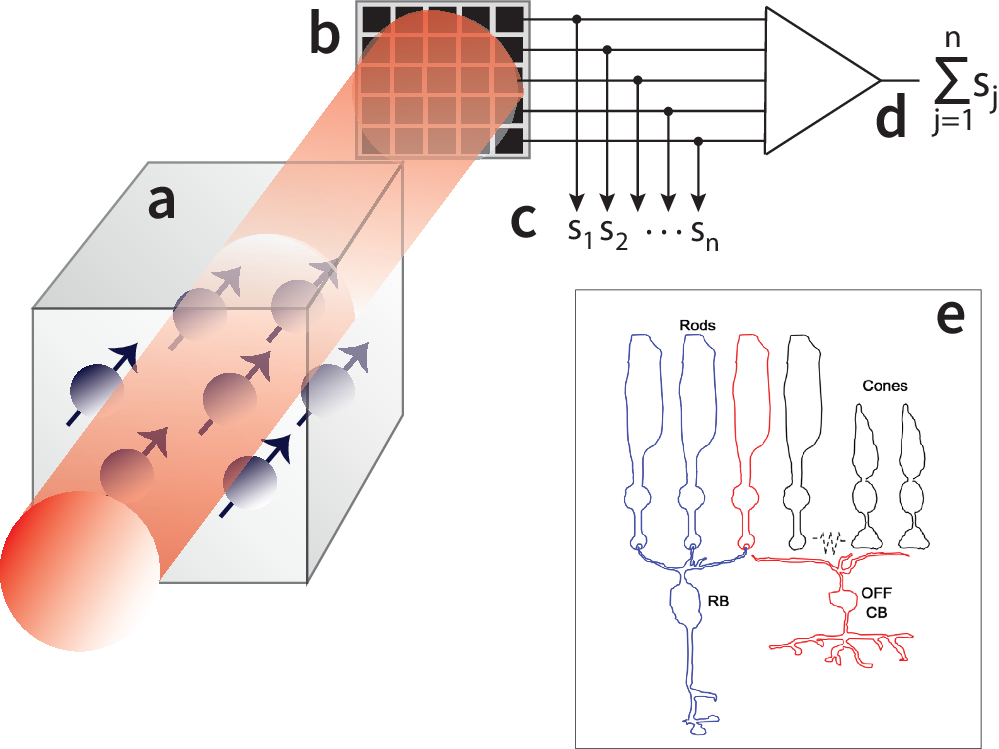}
\caption{(a) Sensitive volume in atomic vapor magnetometers. An ensemble of spins in an atomic vapor is enclosed in a cell and probed by a laser, which is detected by (b) a photodiode array. (c) If one measures the individual signals, $s_1$, $s_2$, ..., $s_n$, produced by the array, then the sensitive volume is determined by the area of one photosensitive element and the laser path length in the vapor. (d) If one integrates the $n$ signals produced by the array, the sensitive volume is determined by the whole area of the photodiode array and the path length, essentially the total volume of the vapor (if the laser waist covers the whole cell).} 
\label{as}
\end{center}
\end{figure}

An analogy with optical pumping magnetometers \cite{BR} will facilitate this discussion. Such magnetometers, schematically shown in Fig. \ref{as}a, involve atomic spins in the vapor phase, with the atomic vapor enclosed in a glass cell. The atoms' spin state, conveying information about the magnetic field, is probed by a laser. One can use a photodiode detecting the whole laser beam, alternatively, one can use a photodiode array (Fig. \ref{as}b). The array consists of $n$ photosensitive elements. If one detects the individual signals $s_1$, $s_2$, ..., $s_n$, then each signal conveys information about the magnetic field experienced by the atoms inside the volume defined by the area of one element and the laser path length through the vapor (Fig. \ref{as}c). If, on the other hand, the $n$ individual signals are integrated (Fig. \ref{as}d), information about the magnetic field within the aforementioned elementary volumes is lost, and the sensitive volume is now defined by the laser path length in the vapor and the whole area of the photodiode array (essentially, this is the total vapor volume, when the laser beam waist covers the whole cell). We note that this physical picture assumes the presence of a buffer gas used to hinder atomic collisions with the walls, in particular, with a buffer gas pressure such that the diffusion time through the aforementioned elementary volumes is not shorter than the measurement time.

Biological systems have devised integration mechanisms in analogy with the aforementioned photodiode array. For example, rod cells in the retina are integrated by bipolar cells, the receptive field of the latter resulting from a sort of superposition of the receptive fields of the former \cite{retina}. It is thus not surprising that radical-pair magnetoreception also involves the integrated response of a number of magnetoreceptive elements, which contribute to the biochemical signal conveying a magnetic field change ($B\rightarrow B+\delta B$) to the animal's brain. This is necessitated by the fact that one molecule alone cannot provide the required magnetic sensitivity, which finally is obtained by a statistical averaging of the responses of $N_{\rm cry}$ radical-pairs residing in $N_{\rm cry}$ Cry proteins. In the previous subsection the volume attributed to those $N_{\rm cry}$ elementary sensors was $N_{\rm cry}V_{\rm rp}$. This choice should be accompanied by a disclaimer.

In the derivation of the ERL presented in \cite{KominisERL}, it is shown that the information flow between \enquote{system} and \enquote{meter} plays a crucial role. In quantum measurements, two parties are involved in the measurement process, the \enquote{system} e.g. the spin, and another physical object interacting with the system, called the  \enquote{meter}. Due to their interaction, information about the system's state is encoded in the meter. The information flow between system and meter takes place inside the sensitive volume during the measurement time. The measurement time is determined by spin relaxation effects. {\it Thus the sensor volume should be the physical volume where those relaxation effects terminating the radical-pair reaction are unfolding}.
One can readily argue that such spin relaxation effects (related to electron transfer, molecular rotation, exchange and dipolar interactions between the two unpaired electrons of the radical-pair \cite{Efimova}) are localized within the radical-pair volume $V_{\rm rp}$. The reaction products could diffuse elsewhere, and they might trigger many more information transfer processes at larger spatial scales, but fundamentally, the termination of the measurement is the end of the reaction, taking place within the volume $V_{\rm rp}$ defined by ${\rm FAD^{\bullet -}Tr^{\bullet +}}$. 

However, one could also imagine \enquote{non-local} interactions at the spatial scale of the whole Cry that could affect spin dynamics of the ${\rm FAD^{\bullet -}Tr^{\bullet +}}$ radical-pair. For example, one could consider a free radical at the \enquote{outskirts} of Cry, i.e. a distance of about 10 nm away from the ${\rm FAD^{\bullet -}Tr^{\bullet +}}$ pair. The dipolar magnetic field of the free radical would produce at the site of ${\rm FAD^{\bullet -}Tr^{\bullet +}}$ a magnetic field of about $10~\upmu{\rm T}$, comparable to earth's field. This seems hard to reconcile with a working magnetoreceptor, on the other hand, the cumulative effect of several such free radicals could be suppressed, and lead to non-local relaxation consistent with the magnetic sensitivity of the sensor.  Another possibility that has been touched upon in the literature is the role of hyperfine relaxation from the protein's proton spins \cite{Walters}, which effect is not yet broadly understood. 

Thus, the disclaimer is that while we use $V=V_{\rm rp}$, a more refined understanding of spin-relaxation in such reactions will not only constrain $\tilde{\tau}$, but might also lead to $V>V_{\rm rp}$. Either possibility will translate into a different $N_{\rm cry}^0$. But in any case, with $N_{\rm cry}$ radical-pairs collectively producing the magnetic sensing response, the sensor volume scales as $N_{\rm cry}$, meaning that the magnetic sensitivity $\delta B$ scales as $1/\sqrt{N_{\rm cry}}$. This is the standard statistical gain over many independent repetitions of a measurement \cite{BR}, usually referred to as standard quantum limit (see also \cite{qb9}).

Similarly (to the extent that radical-pair relaxation is non-local), in atomic sensors the sensitive volume is not given by the atomic volume (about $1~ {\rm \mathring{A}}^3$) times the number of atoms in the vapor, but instead \cite{MitchellRMP}, it is defined by the volume occupied by the whole vapor. On the one hand, for buffer gas pressures as large as e.g. 5 atm, the diffusion constant of alkali atoms in e.g. nitrogen is $D\approx 0.06~{\rm cm^2/s}$, thus within the spin coherence time of about $T_2\approx 10~{\rm ms}$ the atom samples a volume of $(DT_2)^{3/2}\approx 10^{-11}~{\rm m^3}\gg 1~ {\rm \mathring{A}}^3$. Furthermore, in atomic vapors the meter is another atom of the vapor itself, with every atom experiencing many collisions with other atoms and exchanging information about each others' spin state, while relaxing the spin within the time $T_2$. It is thus all kinds of binary collisions within the whole vapor that drive these effects. The probing laser cannot resolve local effects at the scale of binary atomic collisions, the information on which is spread over the volume of the whole vapor, hence it is this volume that defines the sensitive volume. 

In the case of CPF in solution, the typical distance between molecules is 25 nm, while the size of CPF is about 8 nm. We here expect solvation effects in electron-transfer dynamics to make non-local spin relaxation effects non-neglibigle. The probing laser does not spatially resolve spin relaxation effects at the scale of one molecule or its solvation cage, thus we defined the sensitive volume with the total volume of the solution probed by the laser. This is equivalent to using the cage volume of $(25~{\rm nm})^3$ per molecule and multiplying by the number of molecules.
\section{Magnetite Mechanism}
The magnetite mechanism is based on the idea of a small permanent magnet moving like a compass needle due to its interaction with the geomagnetic field \cite{magn2}. Magnetotactic bacteria use magnetite or greigite particles for navigating in the aquatic environment \cite{bacteria1,bacteria2,bacteria3,bacteria4,bacteria5}. It was the presence of ferrimagnetic materials in numerous magnetoreceptive animal species \cite{Gould,Liang,Eder,Diebel,Walker,Beason} that inspired the magnetite mechanism \cite{Kirschvink1981,Kirschvink1989,Begall}.

\begin{figure}[t!]
\begin{center}
\includegraphics[width=8.5 cm]{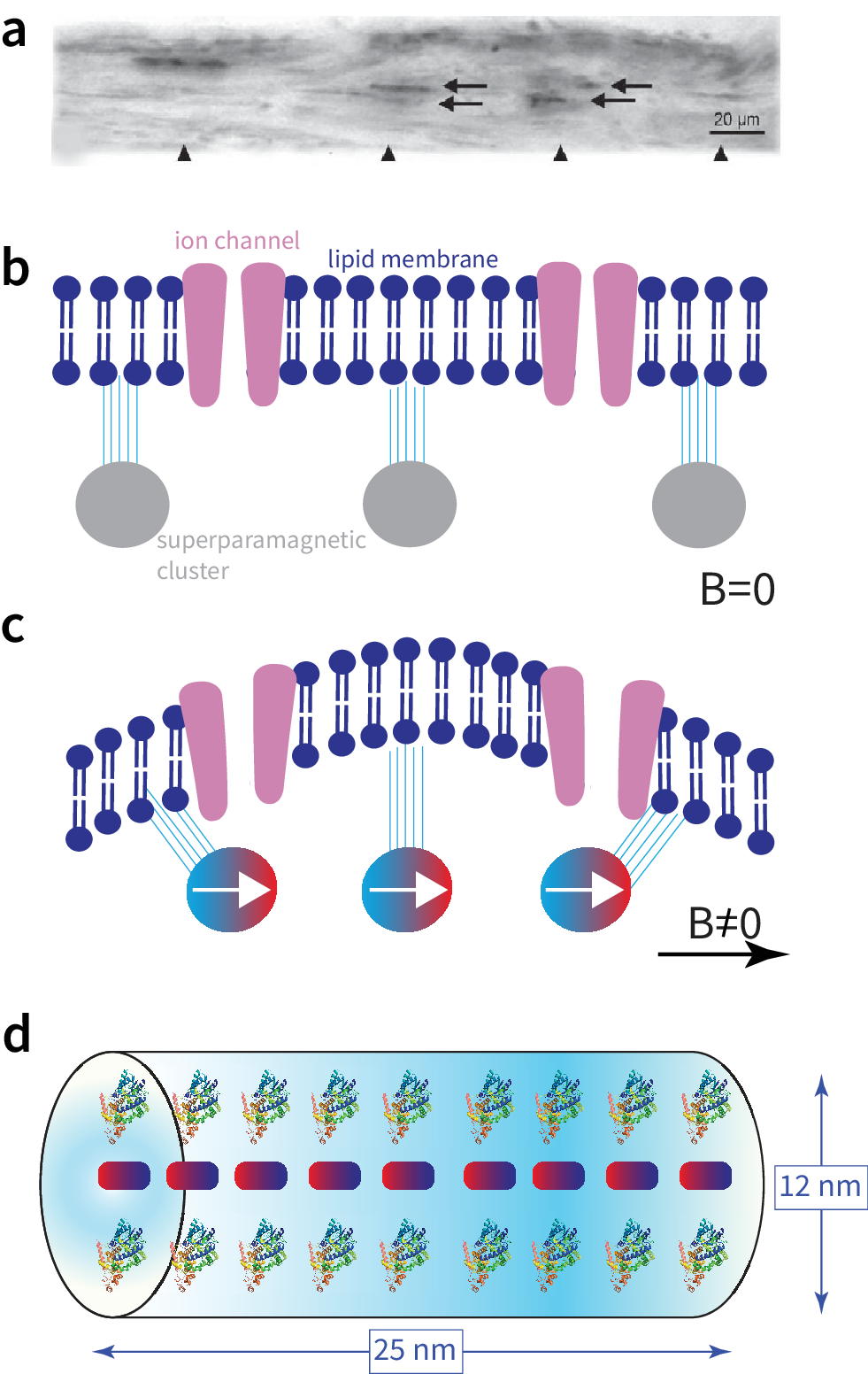}
\caption{(a) Axon bundle with several iron containing dendrites (figure reproduced from \cite{Falkenberg} under the Creative Commons Attribution License). (b) Magnetoreception model based on a chain of superparamagnetic clusters attached to a membrane \cite{Davila}. (c) When the magnetic field is non-zero and points along the chain, the clusters get magnetized along the same direction and attract each other, deforming the membrane and conveying the magnetic field change by changing ion-channel currents. If the magnetic field points perpendicular to the chain (not shown here), the clusters repel each other. (d) Geometry of the MagR-Cry protein complex as derived from the measurements presented in \cite{Xie1}.}
\label{magn}
\end{center}
\end{figure}

Since such materials were located in the upper beak, in particular in the ophthalmic nerve (Fig. \ref{magn}a) of birds \cite{Fleissner,Falkenberg,Heyers,Wiltschko2013}, the general physical process that emerged assumes motion of the magnetic material caused by earth's field, and detection of this motion by neurophysiological means \cite{Winklhofer1999,Fleissner2007,Greiner2007,Winklhofer2010}.  

We will here consider a representative such model. The authors in \cite{Davila} consider superparamagnetic clusters of size 1 ${\rm \upmu m}$, in particular a chain of 20 such clusters separated by about 1.6 ${\rm \upmu m}$. Thus, the volume of the sensor is  $V\approx 20\times 1.6~{\rm \upmu m}\times \pi (0.5~{\rm \upmu m})^2\approx 3\times 10^{-17}~{\rm m}^3$. The superparamagnetic particles are supposed to be attached to a neuron's membrane, as shown in Fig. \ref{magn}b depicting the geometry at zero applied field. For a non-zero magnetic field along the chain, the particles are magnetized as shown in Fig. \ref{magn}c, pulling each other and deforming the membrane \cite{Torbati}. This deformation could alter the open/closed probability of ion channels, and thus convey the magnetic field change to the brain via action potentials. The authors estimate a force acting between the clusters of order $F_m=10^{-13}\chi^2~{\rm N}$, where $\chi$ is the susceptibility of the superparamagnetic material. From relevant measurements \cite{Grob} it follows that $\chi\approx 1$, thus $F_m\approx 10^{-13}~{\rm N}$.

To evaluate the measurement time $\tau$ we neglect the elastic properties of the filaments attaching the clusters to the membrane and consider just two forces acting on the spherical cluster, the magnetic force $F_m$ and the drag force of the cytoplasm $F_d$. The latter is given by $F_d=6\pi\eta_crv$, where $\eta_c\approx 3.5\times 10^{-3}~{\rm Ns/m^2}$ is the cytoplasm viscosity, known to be 5 times that of water \cite{Ashmore}, $v$ the sphere's velocity, and $r$ its radius. The sphere, of mass density about $\rho_s=5~{\rm g/cm^3}$, reaches its terminal velocity, $v=F_m/6\pi\eta_c r\approx 3~{\rm \upmu m/s}$, within time $2\rho_sr^2/9\eta_c\approx 80~{\rm ns}$.

We now assume that the measurement time is given by the time it takes for the sphere to travel a distance equal to the ion-channel's width of order 1 nm \cite{Coates}. Using the value of the terminal velocity we find $\tau\approx 0.3~{\rm ms}$. In fact, this timescale also reflects the typical frequency of elastic vibrations of the membrane. Indeed, the authors in \cite{Tamayo} measure a fundamental flexural frequency of about 10 kHz for a cell having diameter $10~{\rm \upmu m}$. If the surface of the deformed membrane considered herein is determined by the length of the chain, i.e. $32~\upmu {\rm m}\times 32~\upmu {\rm m}$, and considering that the frequency is inversely proportional to the square root of the membrane's mass (and thus its surface area), we find a flexural mode of 5.5 kHz for our case, translating to $\tau\approx 0.2~{\rm ms}$. We thus set $\tau=0.3~{\rm ms}$. 

Putting it altogether, it follows that ${\rm ER}=3\times 10^{6}\hbar$. We note that works like \cite{Jandacka,Winklhofer2022} questioned aspects of magnetite-based magnetoreception. From the perspective of the ERL, the particular mechanism examined here is physically allowed, nevertheless, it should be obvious that the ERL does not prove the mechanism is actually realized as thought at the physiological level. Moreover, the ERL demonstrates that this mechanism requires just one unit of this sensor (i.e. just one chain of superparamagnetic clusters) for obtaining the magnetic sensitivity $\delta B$ within the time $\tau$. 

Overall, the values for $\widetilde{\delta B}=\delta B\sqrt{\tau}$ and the spatial dimension $L$ that we will use for magnetite magnetoreception in our global results of Sec. VII  are $\widetilde{\delta B}\approx 5~{\rm nT/\sqrt{Hz}}$ and $L=V^{1/3}=3~\upmu {\rm m}$.
\section{MagR}
The authors in \cite{Xie1,Xie2,Xie3,Xie4,Xie5,Xie6,Xie7,Xie8,Xie9} used genome-wide screening and a tour-de-force experimental validation to discover an iron-sulphur cluster protein, called MagR, which forms a rod-like complex with cryptochromes. The MagR complex has been studied further \cite{Zhou2016,Jiang2017,Chang2017}, and is already used in magnetogenetic applications \cite{Xue2020,Kang2021}. The geometry of the MagR complex is shown in Fig. \ref{magn}d. It has a length of 25 nm and diameter of 12 nm, thus volume $V=3\times 10^{-24}~{\rm m^3}$. The exact workings of this complex are not yet understood. It seems reasonable to assume that Cry proteins work synergistically with MagR, in some sort of synthesis of the two mechanisms considered previously.  We here make this working assumption. 

Since the biophysics of the MagR complex remains to be unravelled, a hand-waving approach for estimating the measurement time $\tau$ would be to use the geometric mean of the values of the two previous mechanisms, Cry ($10~\upmu{\rm s}$), and magnetite ($300~\upmu{\rm s}$). Thus $\tau\approx 60~\upmu{\rm s}$. A more quantitative approach could follow if we use the scenario \cite{Xie2} that the iron-sulphur complex relates to the electron transfer processes in Cry. From the geometry in Fig. \ref{magn}d we estimate the distance between the iron-sulphur complex and the Cry to be somewhere between 3 nm and 4 nm. At such long distances, we find \cite{et1,et2} electron transfer rates that range from $10^3~{\rm s}^{-1}$ to $10^5~{\rm s}^{-1}$. We thus set $\tau=100~\upmu{\rm s}$. Towards a more general treatment we introduce the dimensionless factor $\tilde{\tau}={\tau\over {100~\upmu {\rm s}}}$, and let $\tau$ vary between  $10~\upmu {\rm s}$ and $1000 ~\upmu {\rm s}$, thus $\tilde{\tau}$ ranges from 0.1 to 10.

The resulting ER is $0.1\tilde{\tau}\hbar$, meaning that $N_{\rm MagR}^0(\tilde{\tau})=1/(0.1\tilde{\tau})=10/\tilde{\tau}$ complexes need to be integrated for the mechanism to hit the ERL bound. It is clear that due to the much larger volume of the complex compared to the case of radical-pair magnetoreception, fewer units are required to be right on the ERL, thus fewer Cry proteins overall. Equivalently, for a limited number of complex constituents, $\delta B$ decreases when the sensitive volume of the complex doing the sensing increases. We could imagine that this might incur some evolutionary advantage of this particular design, however, it is rather early to make relevant statements, given the limited understanding of this magnetoreceptor.
\begin{figure*}[ht]
\begin{center}
\includegraphics[width=17.8 cm]{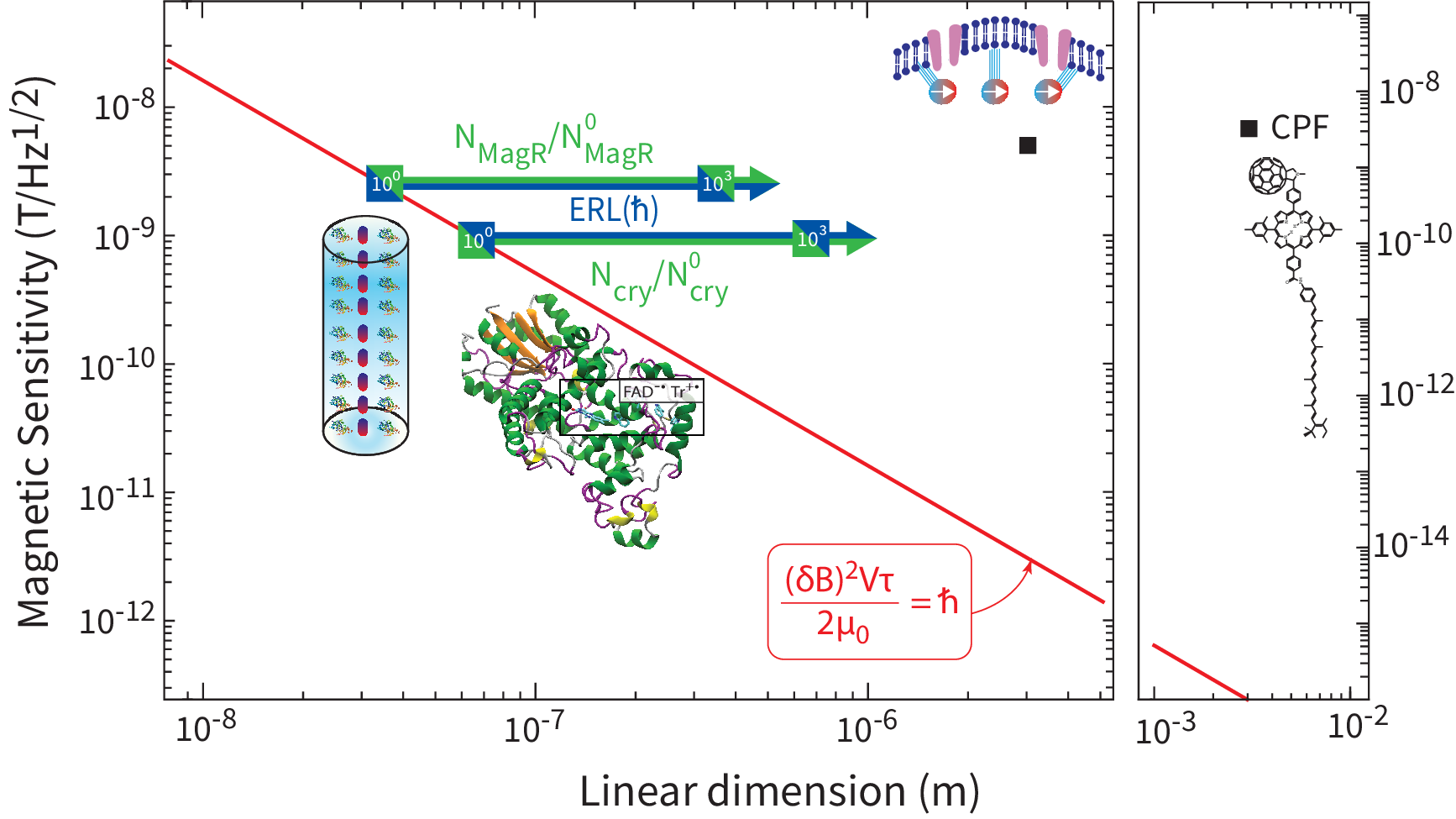}
\caption{Magnetic sensitivity $\widetilde{\delta B}=\delta B\sqrt{\tau}$ as a function of $L$, the linear dimension of the sensor, for several mechanisms considered in animal magnetoreception. The red solid line defines the energy resolution limit given by $\widetilde{\delta B}=\sqrt{2\mu_0\hbar/L^3}$. (a) Cryptochrome magnetoreception based on the FAD-Tr radical-pair sits right at the limit when $2\times 10^5$ cryptochromes contribute to a magnetic sensitivity of $\delta B=0.3~\upmu{\rm T}$ within time $\tau=10~\upmu{\rm s}$. The shaded region accounts for a range of $\tau$ given in terms of the dimensionless parameter $\tilde{\tau}$, i.e. $\tau=10\tilde{\tau}~\upmu{\rm s}$. The number of cryptochromes required to hit the ERL limit of $\hbar$ is a function of $\tilde{\tau}$, i.e. $N_{\rm cry}^0(\tilde{\tau})=2\times 10^5/\tilde{\tau}$. If the number of contributing cryptochromes is e.g. $10^3$ times larger than $N_{\rm cry}^0(\tilde{\tau})$, the total volume of the radical-pair sensor, and thus the ER, also increase by a factor $10^3$, while the linear dimension increases by 10. This is depicted by the green/blue arrows pointing towards increasing linear dimension. The blue arrow shows the ER in units of $\hbar$, and the green arrow the ratio $N_{\rm cry}/N_{\rm cry}^0(\tilde{\tau})$. (b) Similar comments apply to the MagR sensor, shown to hit the ERL when $N_{\rm MagR}^0(\tilde{\tau})=10/\tilde{\tau}$ MagR-protein complexes contribute to detecting the same $\delta B$ within $\tau=100\tilde{\tau}~\upmu{\rm s}$, where again $\tilde{\tau}$ is a dimensionless parameter. For the magnetite mechanism involving a chain of superparamagnetic clusters it is $\widetilde{\delta B}=5~{\rm nT/\sqrt{Hz}}$ for a linear dimension of $3~\upmu{\rm m}$. (c) The laser spectroscopic measurement of CPF takes place in a solution of 3 mm linear dimension, with the corresponding sensitivity being $\widetilde{\delta B}=2.5~{\rm nT/\sqrt{Hz}}$. Picture of cryptochrome protein adapted from \cite{Frederiksen} under the Creative Commons Attribution License.} 
\label{global}
\end{center}
\end{figure*}

Overall, the values for $\widetilde{\delta B}=\delta B\sqrt{\tau}$ and the spatial dimension $L$ that we will use for MagR magnetoreception in our global results of Sec. VII are $\widetilde{\delta B}\approx 3\sqrt{\tilde{\tau}}~{\rm nT/\sqrt{Hz}}$ and $L=(N_{\rm MagR}V)^{1/3}\geq 31/\tilde{\tau}^{1/3}~{\rm nm}$, where $N_{\rm MagR}$ is the number of MagR complexes contributing to the estimate of the magnetic field, and necessarily, it is $N_{\rm MagR}\geq N_{\rm MagR}^0(\tilde{\tau})$.
\section{Global ER MAP for biological magnetoreception}
In Fig. \ref{global} we present the global map of the spectral magnetic sensitivity given by $\widetilde{\delta B}=\delta B\sqrt{\tau}$ as a function of the sensor's linear dimension $L$ for all three sensing mechanisms we have considered. We remind the reader that we  used a common value for the magnetic sensitivity, $\delta B=0.3~\upmu{\rm T}$. The parameters that differ for each sensing mechanism are the measurement time $\tau$ and the linear dimension $L$. From the global map we can draw a number of conclusions. 

From the perspective of the ERL, the radical-pair magnetoreception can indeed be \enquote{quantum}, if the number $N_{\rm cry}$ of Cry proteins participating in the sensing equals $N_{\rm cry}^0(\tilde{\tau})=2\times 10^5/\tilde{\tau}$, which number depends on the specific value of $\delta B$ in vivo, the parameter $\tilde{\tau}$ quantifying spin relaxation time, and the sensor's volume. For $N_{\rm cry}>N_{\rm cry}^0(\tilde{\tau})$, the energy resolution grows beyond $\hbar$, and the \enquote{quantumness} is reduced. Similar comments apply to the MagR sensor, with the relevant minimum number of contributing complexes being $N_{\rm MagR}^0(\tilde{\tau})=10/\tilde{\tau}$. 

The magnetite mechanism, and even more so the CPF measurement appear to be more \enquote{classical}, which is probably not surprising. Nevertheless, the ER of the magnetite mechanism stands at the level of $10^6\hbar$, which is not too far from the quantum limit. As interactions of magnetized materials with the magnetic field are at their core quantum, the prospect of pushing the magnetite mechanism closer to the quantum limit with biomimetic engineering seems reasonable.
\section{Conclusions}
Magnetometry is an increasingly active field of quantum technology, with several magnetic sensing technologies competing in terms of sensitivity and applications in fundamental and applied science. The energy resolution is a concept unifying numerous magnetic sensing technologies. Nature has also developed magnetometers, since the geomagnetic field offers a rather stable navigation cue. Here we made the connection between the energy resolution limit, a concept developed by the quantum sensing community, with biological magnetoreception. We thus arrived at a model-independent consistency check, which informs the workings of several mechanisms aspiring to explain animal magnetoreception. The energy resolution of each magnetoreception mechanism not only provides information about uncertain system parameters, but also illuminates the room for improvement, were one to design a biomimetic sensor. Our considerations thus offer a new perspective on quantum biology, synthesizing quantum sensing with biological sensing, and highlighting the future potential of such synthesis. 

In particular, the most obvious utility of the energy resolution limit is to extract information about unknown system parameters. For example, given the magnetic sensitivity and the measurement time of the Cry and MagR magnetoreceptors, we extract for the first time the minimum number of magnetoreceptors required for sensing, since it is this minimum number that is required to satisfy the ERL bound. Conversely, the fact that a given magnetic sensing model might satisfy the ERL, as for example the magnetite model does, is not proof that the model is correct, but merely shows it is a physically viable model.

The results presented here will evolve together with a more precise understanding of the relevant parameters expected from future experiments in vivo and in vitro. For example, we used a common value for the magnetic sensitivity $\delta B$ applicable to all magnetoreception mechanisms and animal species. This might be better specified in the future, however the utility of the ERL will be the same, still helping to constrain the sensor's three parameters involving sensitivity, volume and measurement time. 

From the quantum engineering perspective, the ERL informs about the extent that a given naturally appearing magnetoreception mechanism can be improved if one develops biomimetic sensors. This work can thus support quantum biotechnology inspired by Nature.
\acknowledgements
We kindly acknowledge helpful comments by Kevin Henbest, Peter Hore and Michael Winklhofer.

\end{document}